\documentclass[sigconf,accept]{acmart}

\AtBeginDocument{%
  \providecommand\BibTeX{{%
    \normalfont B\kern-0.5em{\scshape i\kern-0.25em b}\kern-0.8em\TeX}}}
\copyrightyear{2022} 
\acmYear{2022} 
\setcopyright{acmlicensed}\acmConference[ICSE-NIER'22]{New Ideas and Emerging Results }{May 21--29, 2022}{Pittsburgh, PA, USA}
\acmBooktitle{New Ideas and Emerging Results (ICSE-NIER'22), May 21--29, 2022, Pittsburgh, PA, USA}
\acmPrice{15.00}
\acmDOI{10.1145/3510455.3512773}
\acmISBN{978-1-4503-9224-2/22/05}

\usepackage[english]{babel}
\usepackage[utf8x]{inputenc}
\usepackage[colorinlistoftodos]{todonotes}
\usepackage{framed}
\usepackage{mdframed}
\begin{document}

\title{\toolname: A Context-Aware Code Analysis Tool}

\author{Jai Kannan}
\email{jai.kannan@deakin.edu.au}
\author{Scott Barnett}
\email{scott.barnett@deakin.edu.au}
\affiliation{%
  \department{Applied Artificial Intelligence Inst.}
  \institution{Deakin University}
  \city{Geelong}
  \state{Victoria}
  \country{Australia}
  }



\author{Luís Cruz}
\email{l.cruz@tudelft.nl}
\affiliation{%
  \institution{Delft University of Technology}
  \city{Delft}
  \country{Netherlands}
  }

\author{Anj Simmons}
\author{Akash Agarwal}
\email{a.simmons@deakin.edu.au}
\email{a.agarwal@deakin.edu.au}

\affiliation{%
  \department{Applied Artificial Intelligence Inst.}
  \institution{Deakin University}
  \city{Geelong}
  \state{Victoria}
  \country{Australia}
  }

\begin{abstract}
Meeting the rise of industry demand to incorporate machine learning (ML) components into software systems requires interdisciplinary teams contributing to a shared code base. To maintain consistency, reduce defects and ensure maintainability, developers use code analysis tools to aid them in identifying defects and maintaining standards. With the inclusion of machine learning, tools must account for the cultural differences within the teams which manifests as multiple programming languages, and conflicting definitions and objectives. Existing tools fail to identify these cultural differences and are geared towards software engineering which reduces their adoption in ML projects. In our approach we attempt to resolve this problem by exploring the use of context which includes i) purpose of the source code, ii) technical domain, iii) problem domain, iv) team norms, v) operational environment, and vi) development lifecycle stage to provide contextualised error reporting for code analysis. To demonstrate our approach, we adapt Pylint as an example and apply a set of contextual transformations to the linting results based on the domain of individual project files under analysis. This allows for contextualised and meaningful error reporting for the end user.
\end{abstract}



\newcommand{\toolname}{MLSmellHound}

\maketitle

\section{Introduction}
\label{sec:introduction}

Code analysis tools reduce maintenance costs of a system through the early detection and prevention of defects \cite{Dai2017,Chen2016,Marcilio2020,Paul2021,Kaur2020,Mokhov2015}. These tools also support the identification and refactoring of code smells \cite{Lavazza2021,Cairo2018} and reducing ML specific technical debt. By integrating code analysis tools into CI/CD pipelines modifications to code can be tested to improve i) security \cite{Bafatakis2019}, ii) readability \cite{Posnett2011}, and iii) consistency \cite{Zou2019}. 

Machine learning (ML) applications stand to benefit from code analysis by i) identifying ML specific technical debt ~\cite{Sculley2015}, ii) catching defects in models and data and iii) mitigating ML unique failure modes~\cite{Baier2020,Kumar2019}. These unique failure modes contribute to eroded trust and reputation damage as demonstrated by Amazon's sexist recruiting tool \cite{reuters2018} or failure of the Thistle F.C. ball tracking system \cite{verge2020}. 
ML systems also have increased maintenance costs and effort to manage infrastructure for data and model management~\cite{Sculley2015}. The additional infrastructure and dependence on data increases area where defects can occur. Code analysis tools are a promising approach for improving the robustness of ML applications.  

However, code analysis tools suffer from high false positive rates when analysing ML applications~\cite{VanOort2021a}. The impact of high false positive rates is poor adoption and wasted time. False positives are perceived as a challenge \cite{Tomasdottir2020, Tomasdottir2017}. Calibration is required to balance false negatives (missed defects) and false positives (no defect present). Development of ML applications utilises multiple languages and mixed teams of data scientists (specialists in development and application of ML algorithms) working alongside software engineers to integrate the algorithms with the surrounding infrastructure. As such, defects in ML applications require a combination of data science and software engineering expertise to resolve~\cite{haakman2020ai}. Thus, code analysis tools operating on ML applications must account for i) multiple programming languages, ii) conflicting definition and prioritisation of defects in interdisciplinary teams, iii) ML specific failure modes, and iv) provide contextualised error reporting to provide explanations in the context of ML to reduce false positives.


Existing code analysis tools for ML have focused on data linters \cite{Hynes}, and tensor shapes \cite{lagouvardos2020}. Code analysis tools  focus on analysing Python, a popular language for ML \cite{Simmons2020} or identifying general code smells \cite{Gulabovska2019}. However, existing approaches are not sufficient to provide actionable explanations for mixed development teams and ignore the technical domain of ML \cite{Barnett2018}. For example, during exploratory data analysis, code quality is sacrificed for speed of experimentation and discovery of insights; here the activity influences the definition of code quality. This paper proposes inclusion of context from the technical domain, software artefact and environment to improve the precision of code analysis tools through contextualised feedback.

In this paper we present a research agenda for \toolname{}, a context-aware tool that improves the precision of code analysis tools. The goal of our research is to improve the maintainability and robustness of ML applications through context-aware code analysis tools. We hypothesize that context can improve the precision of code analysis tools and propose the following research questions:
\begin{itemize}
    \item \textbf{RQ1:} What context can be mined from ML projects? 
    \item \textbf{RQ2:} How can context transformations be applied to prioritize ML smells?
    \item \textbf{RQ3:} Which software metrics can be used with context to predict ML smells? 
\end{itemize}
Our novel approach is inspired by ideas from model driven engineering that uses  i) a context metamodel, ii) a set of context transformations, and iii) context checkers to improve code analysis for ML.  

Motivated by the work in \citeauthor{Simmons2020}~\cite{Simmons2020}, we provide a demonstrator of \toolname{} on Python coding conventions. We apply our approach to an existing code analysis tool, Pylint. Our tool uses the purpose of the source code to selectively apply and customise linting errors. 
Contributions arising from this paper include: \textbf{1)} an approach for leveraging \textbf{context} to reduce false positive rates in code analysis tools, and \textbf{2)} a demonstrator of \textbf{\toolname{}} for leveraging context in coding conventions for ML applications.

\section{Motivation}
\label{sec:motivation}

To motivate our research we use coding conventions as an example.

\begin{figure}[h]
  \includegraphics[width=0.8\linewidth]{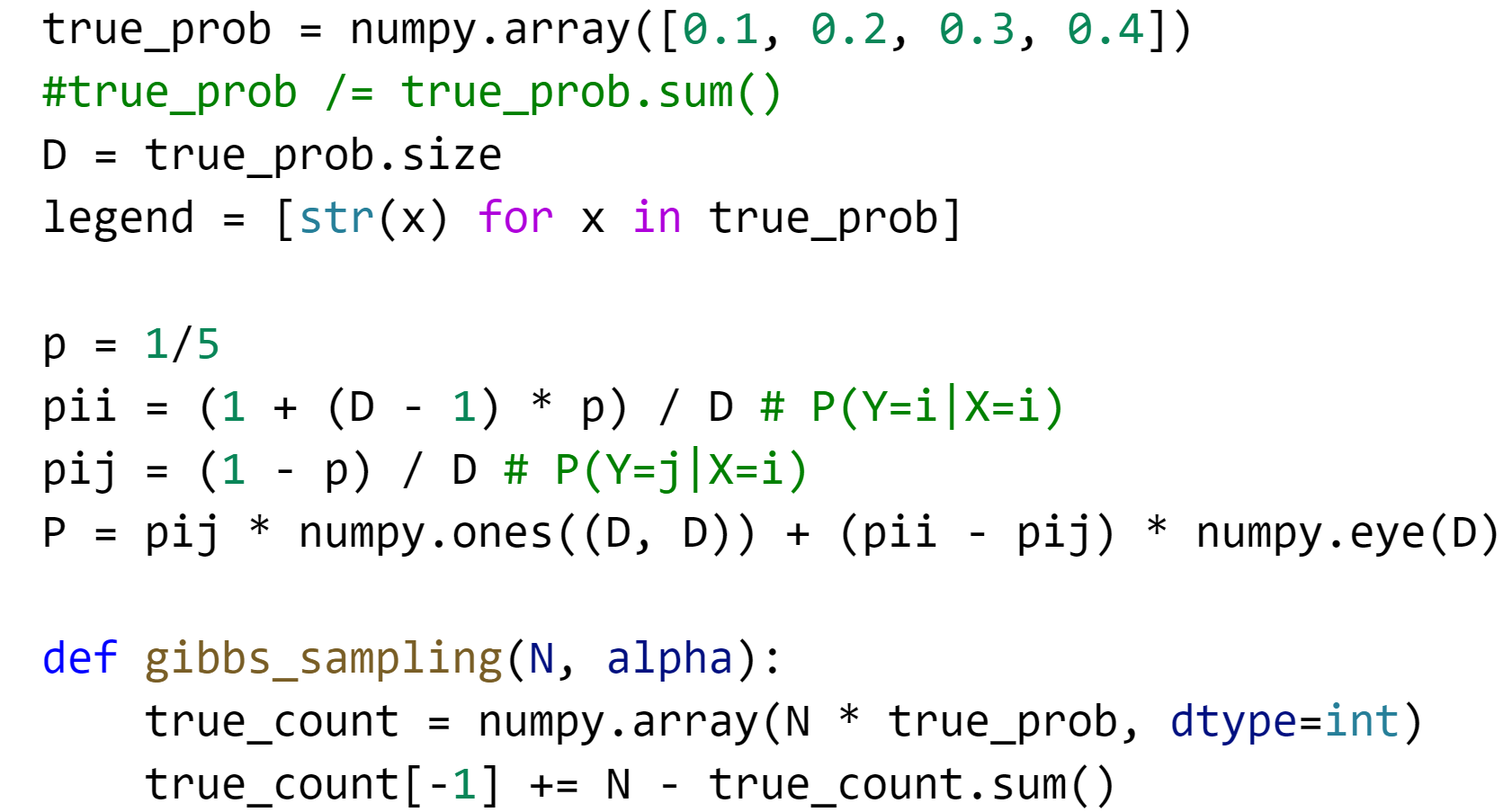}
  \caption[Code snippet]{Code snippet from an open-source project\footnotemark{} violating Python's naming conventions}
  \label{example_source_code}
\end{figure}

Consider Dave, a data scientist, working with a software engineer, Tammy. Dave and Tammy are tasked with creating a fraud prediction model. Dave finds a research paper providing a solution that best fits the description of the problem. Dave implements the solution in Python and the widely available ML frameworks \cite{Braiek2018}. When Dave issues a pull request, the pull request is reviewed by Tammy. During the review Tammy identifies that the code violates their organisation's coding standards. \autoref{example_source_code} shows an example of Dave's code.
\footnotetext{https://github.com/shuyo/iir.git}

The organisation follows PEP8 coding standards, that states that variable names such as `P' should be lowercase. When asked why Dave neglected to use the organisation's code analysis tool, Dave responds that he prefers keeping the variable names similar to the mathematical notation of the algorithm in the paper he is implementing, for example, capital letters to represent matrices, and indicates that the configuration used by the organisation ignores data science norms resulting in high false positives from his perspective. Dave and Tammy agree there is a need to adjust the organisation's standards to restore team cohesion; however, deliberating a fix diminishes productivity as i) existing linter configurations are calibrated, ii) rule sets defined and agreed upon, and iii) new scripts are implemented as existing tools cannot identify the context of ML code. To help Dave and Tammy we propose a context-aware tool that selectively applies convention rules depending on the code context.

\section{Vision}
\label{sec:vision}

\begin{figure*}[!tb]
    \centering
    \includegraphics[width=0.8\linewidth]{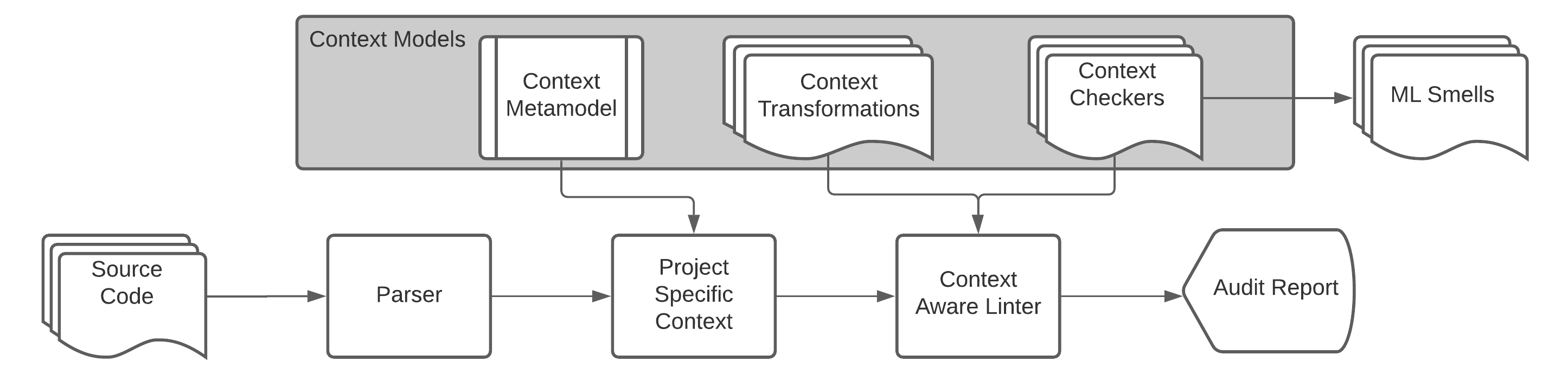}
    \caption{A conceptual overview of the novel context-aware code analysis tool, \toolname{}.}
    \label{fig:system_overview}
\end{figure*}

Our goal is to improve the maintenance aspects and robustness of ML applications by using context-aware code analysis tools. We borrow ideas from model driven engineering (MDE)~\cite{Bezivin2006} to compose a new category of tools for modelling context. These tools are designed to provide better tool support for interdisciplinary teams.Purpose of the code determines the priority and importance of the defect E.G: Experiment/ Deployment To realise our vision we have modelled the problem with concepts from MDE as shown in \autoref{fig:system_overview}. We use the principle of meta-modelling \cite{Baudry2008} to create the context meta-models and context transformations for the tool. Below, we pinpoint the core components of the proposed model.

\subsection{Context Metamodel}
\vspace{0.5em}
\begin{mdframed}
\textbf{RQ1. What context can be mined from ML projects?}
\end{mdframed}
The context metamodel represents properties that comprise the context for an ML application. We plan to investigate multiple facets of context including i) purpose of the source code, ii) technical domain \cite{Barnett2018}, iii) problem domain, iv) team norms, v) operational environment, and vi) development lifecycle stage. The context metamodel will define the relationships between these concepts. The source code is given as input to the system. The system makes use of a \textbf{Parser} (c.f \autoref{fig:system_overview}) to abstract over different source code languages and repositories, and map it to contextual features; for example, extracting file level imports of known ML frameworks for that language to help determine if the incoming code is categorised as ML or non-ML code. A context metamodel is applied to annotate the source code, producing a \textbf{Project Specific Context} which provides a high level representation for informing the context-aware linter.

\subsection{Context Transformations} 
\vspace{0.5em}
\begin{mdframed}
\textbf{RQ2. How can context transformations be applied to prioritise ML smells?}
\end{mdframed}
The context transformations enable \toolname{} to modify the linter based on the context of a project. These transformations include:
\begin{itemize}
    \item Subtraction - context disallows certain rules from applying on a file. 
    \item Addition - context selectively applies new rules to better inform the user of design problems. 
    \item Reprioritisation - context used to re-rank linting results to draw attention to high priority items and demote items that are unlikely to be relevant in the given context.
    \item Remessage - operational context produces a different message to an end-user to improve usability. 
\end{itemize}
Context transformations can be used with existing tools extending our approach to all code analysis tools.

\subsection{Context Checkers}
\vspace{0.5em}
\begin{mdframed}
\textbf{RQ3. Which software metrics can be used with the context to predict ML smells?}
\end{mdframed}
The context checkers are the rules that are used to operate on the software artefacts informed by the project specific context. These checkers are informed by ML smells that are known to cause issues and include metrics for identifying each smell. We hypothesise that the combination of context transformations and a context model will improve the precision of code analysis tools. The \textbf{Context-Aware Linter} (cf. \autoref{fig:system_overview})  is a code analyser performing two functions: i) checking for ML code smells and anti-patterns by applying the relevant set of \textbf{Context Checkers} to each file in the source code; ii) generating the context-transformed linting messages. Using these two functions the \textbf{Audit Report} is generated for the user to evaluate the code.

\section{Why is it new?}
\label{sec:futurework}

\textbf{ML smells are an emerging area of research.} Recent work has investigated i) the prevalence of code smells in ML applications~\cite{VanOort2021a}, ii) deep learning~\cite{Jebnoun2020} code smells, iii) anti-patterns \cite{Muralidhar2021}, and linters for datasets~\cite{Hynes}. However, a gap exists for a code analysis tool that automatically identifies ML-specific smells. Unlike existing tools, our approach also considers the technical domain~\cite{Barnett2018} to provide context-aware recommendations for the code base. New tools are needed to identify suitable metrics that locate ML smells and support the developer in prioritising impact to redirect engineering effort.




\textbf{Code analysis tools ignore cultural differences.} 
Current code analysis tools are geared to support software engineering standards, however not all of these standards may be applicable for ML  software, resulting in an increase in false positives and reduced adoption \cite{VanOort2021a, Tomasdottir2020}. Code analysis tools need to consider the cultural differences existing between disciplines in cross functional teams to improve support for shared development. Code quality in ML software may score poorly against traditional code quality metrics due to experimentation, mathematical vocabulary and the nature of code artefacts. To support these developers, we utilise a set of transformations based on the context of the program to selectively apply linting rules. Our tool prioritises outputs for developers, to support data scientists to build standards-compliant code without hindering their productivity.

\textbf{Novel use of context for improving the precision of code analysis tools.}
Context has been used in software engineering to improve software quality \cite{Allamanis2014,Yu2020}, security \cite{Wen2019}, and to improve development time \cite{Ye2020}. In our approach, we leverage MDE to treat code artefacts as models, to extract metadata to improve code analysis for interdisciplinary teams, and to selectively apply a set of transformations to linting outputs with the goal of reducing false positive rates. Contextual features extracted include \textbf{the purpose of the code} (which we plan to detect by analysing the operations implemented in the code). For example, we detect whether a code file pertains to \textbf{ML} functionality, or is intended as a \textbf{non-ML} module that provides the surrounding infrastructure. To the best of our knowledge, this is the first work to use context to improve code analysis for ML projects.

\section{\toolname{} Demonstrator}
\label{sec:resultsandevaluation}

\begin{figure*}[!tb]
    \centering
    \includegraphics[width=0.95\linewidth]{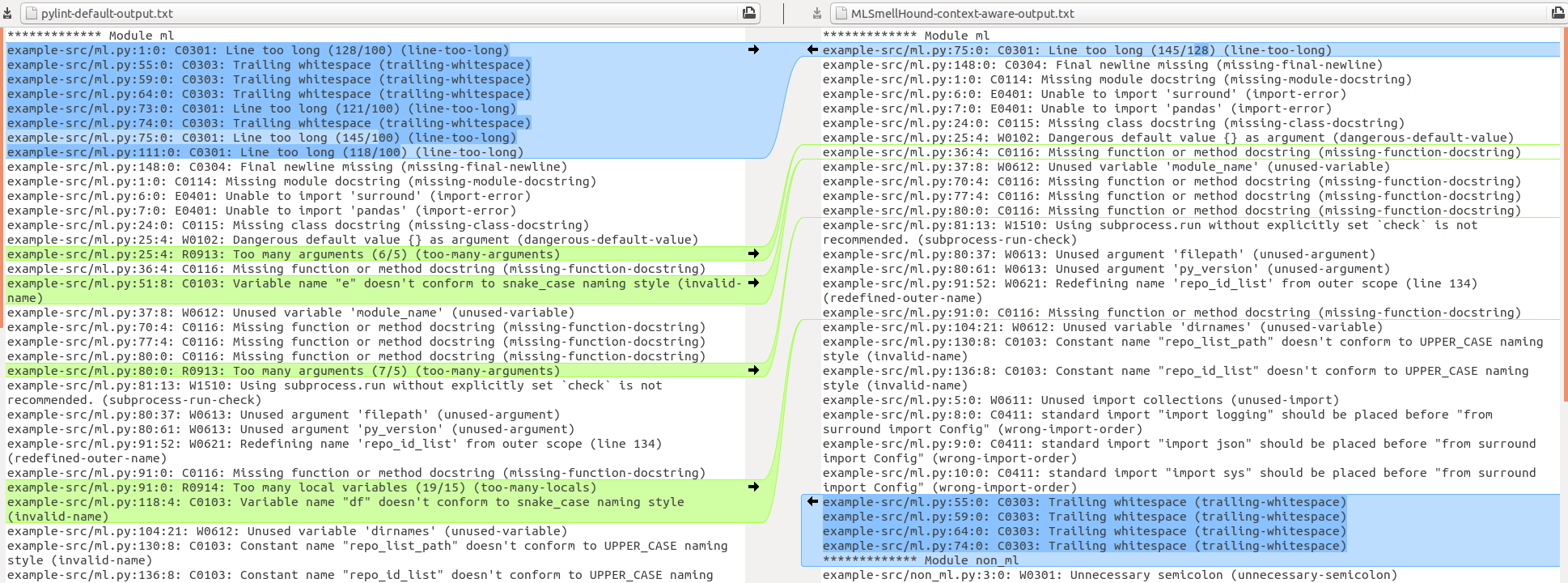}
    \caption{Screenshot showing difference between Pylint default output (left), and transformed ouput generated by \toolname{} (right). Pylint warns about ``e'' and ``df'', whereas \toolname{} accepts these in the context of ML files. The output also shows that trailing-whitespace warnings have been re-ranked to a lower priority for ML files.}
    \label{fig:sample}
\end{figure*}

To demonstrate the approach, we developed an initial prototype of \toolname{} that provides context-aware linting\footnote{Download \toolname{} from: https://github.com/a2i2/ml-smell-hound}. Our tool is based on Pylint, a popular linter for Python projects that is highly configurable but that does not consider context. \textit{The demonstrator shows how existing code analysis tools can be augmented with context using our approach.} For the purposes of demonstration, we use a simplified representation of the purpose of the source code as the context (whether a source file is an \textit{ML} or \textit{non-ML} module), although we will explore finer granularities (e.g. data wrangling, configuration, etc.) and additional dimensions of context (e.g. stage of the team data science process) in the future. To determine this simplified context, we traverse through each of the Python files in a given directory line by line. If we detect the presence of an ML library import, we flag that file as an \textit{ML} module. A sample of the output is presented in \autoref{fig:sample}.


%
The following context transformations are implemented in the tool:
\textbf{Subtraction/Addition:} Depending on the context, we configure Pylint to selectively enable or disable different rules, as well as use adjusted thresholds and patterns. For example, ML code often uses short variable names inspired by mathematical conventions \cite{Simmons2020}, thus we use an altered regular-expression for validating conformance to variable naming conventions when linting ML files. 

\textbf{Remessage:} A common smell detected in ML source code is an excessive number of parameters, which has been speculated to be a result of algorithms with many hyperparameters \cite{Simmons2020}. To ensure the message is understandable to data scientists we provide a simple mechanism for overwriting these messages by mapping Pylint symbols to custom messages tailored to the context.
\textbf{Reprioritisation:} Our tool transforms the linting output by moving message categories that are not relevant for a particular context to the bottom of the generated output. E.g., ML code influenced by Tensorflow often uses non-standard indentation (two spaces instead of four), thus indentation warnings are deprioritised for ML files.

\section{Future Plans}
\label{sec:futurework}

To realise the vision for \toolname{} outlined in \autoref{sec:vision} we have broken our approach down into four phases: i) qualitative evaluation of context for code analysis, ii) catalogue of ML smells and fixes, iii) metrics and context for locating ML smells, and iv) end-user evaluation of \toolname{}. 

\textbf{Qualitative evaluation of context in code analysis.} For further validation, we plan to extend \citeauthor{Simmons2020}~\cite{Simmons2020} by interviewing practitioners to better understand why coding conventions change based on context. The interviews will also be designed to understand the relevant context when building ML applications. 

\textbf{Catalogue of ML smells and fixes.} We plan to conduct a mapping study to mine a catalogue of ML smells and anti-patterns with appropriate solutions. Our goal is to establish if ML anti-patterns map to existing software engineering code smells and analyse their impact on the occurrence of ML failures to create a set of ML-specific smells which map to defects that development teams need to consider while developing ML applications. To improve the completeness of the catalogue we will include both academic and gray literature (i.e. blogs, industry papers, and tutorials). From the catalogue we plan to identify context required for identifying an ML smell and will evaluate our findings with practitioners.  

\textbf{Metrics and context for locating ML smells.} Based on the findings from the previous stage a set of smells will be selected. We will investigate how and where the context can automatically be mined for these smells in a software engineering project. These smells will be integrated into \toolname{} to assess if context can be used to predict other ML smells.  

\textbf{End-user evaluation of \toolname{}.} Our plan is to run \toolname{} against open-source repositories and ask participants to assess the i) accuracy of the reported smells, ii) usability of \toolname{}, and iii) benefit for reducing failure in ML applications. We will also run a questionnaire to evaluate the relevance of context used in the predictions as a form of interpretation of the ML-smell prediction.

\section{Risk}
\label{sec:risk}
Differences in development exist between data scientists and software engineers in software quality and expected outcomes, for example software engineers may value software maintainability while data scientists may value the performance of the model in production. To account for this, we plan to identify these differences and include an approximation of the mental models of each discipline to ensure ML smells and warnings are explained and ranked in a manner that is compatible with the reasoning of that discipline.
 
Another risk for our research is the assumption that dimensions of context influence the relevance of code smells and the appropriate preventative action can be identified and mined from available artefacts. We plan to test this assumption through a controlled study to evaluate the relevance of context with practitioners compared to a control group using a tool that does not take into account context.



\bibliographystyle{ACM-Reference-Format}
\bibliography{references}

\end{document}